\documentclass[%
 reprint,
 showpacs,preprintnumbers,
 amsmath,amssymb,
 aps,
 prb,
]{revtex4-1}
\usepackage{dcolumn}
\usepackage[dvipdfmx]{graphicx}
\usepackage{subfigure}
\usepackage{color}
\usepackage{mathrsfs}

\makeatletter

\begin{document}

\title{Intrinsic spin-valley locking for conducting electrons in metal-semiconductor-metal lateral hetero-structures of $1H$-transition-metal dichalcogenides}

\author{Tetsuro Habe}
\affiliation{Nagamori Institute of Actuators, Kyoto University of Advanced Science, Kyoto 615-8577, Japan}

\date{\today}

\begin{abstract}
Lateral-hetero structures of atomic layered materials alter the electronic properties of pristine crystals and provide a possibility to produce useful monolayer materials.
We reveal that metal-semiconductor-metal lateral-hetero junctions of $1H$-transition-metal dichalcogenides intrinsically possess conducting channels of electrons with spin-valley locking, e.g., gate electrode.
We theoretically investigate the electronic structure and transport properties of the lateral-hetero junctions and show that the hetero-structure produces conducting channels through the $K$ and $K'$ valleys in the semiconducting transition-metal dichalcogenide and restricts the spin of the conducting electrons in each valley due to the valley dependent charge transfer effect.
Moreover, the theoretical investigation shows that the hetero-junction of WSe$_2$ realizes a high transmission probability for valley-spin locked electrons even in a long semiconducting region.
The hetero-junction also provides a useful electronic transport property, a step-like I-V characteristic.
\end{abstract}

\maketitle
\section{Introduction}
Atomic layered materials are solids consisting of atomically thin crystals stacked in layers.
Modern experimental techniques enable one to produce atomically thin materials of the layers by cleaving from the bulk crystal,  chemical vapor deposition,\cite{Lee2012,Bergeron2017}, chalcogenation\cite{Chen2017}, or molecular-beam epitaxy\cite{Moreau2010,Zhan2017,Kazzi2018}.
These thin layers have attracted much attention because of the fascinating electronic structures different from those of the bulk crystals.\cite{Neto2009,Xiao2012}
Moreover, the fabrication techniques to combine different layers into electric or optical systems have been developed in terms of application. \cite{Kang2013,Habe2015,Li2015,Najmzadeh2016,  Son2016,Ullah2017}
These atomically thin systems are expected for the realization of atomic layer electric circuits by combining different atomic layered materials.

Transition-metal dichalcogenides(TMDCs) are solid composed of transition-metal and chalcogen atoms, and include atomic layered materials in the $1H$-structure,  a crystal structure of monolayer.
In $1H$-monolayer crystals, electrons possess two degrees of freedom, spin and valley because of the crystal structure.
The manipulation of spin and valley is one of the attractive purposes in the research field of atomic layered materials.
Since the spin and valley are coupled due to Zeeman-type spin-orbit coupling in the monolayers, these materials are good candidates for these purposes.
\begin{figure}[htbp]
\begin{center}
 \includegraphics[width=76mm]{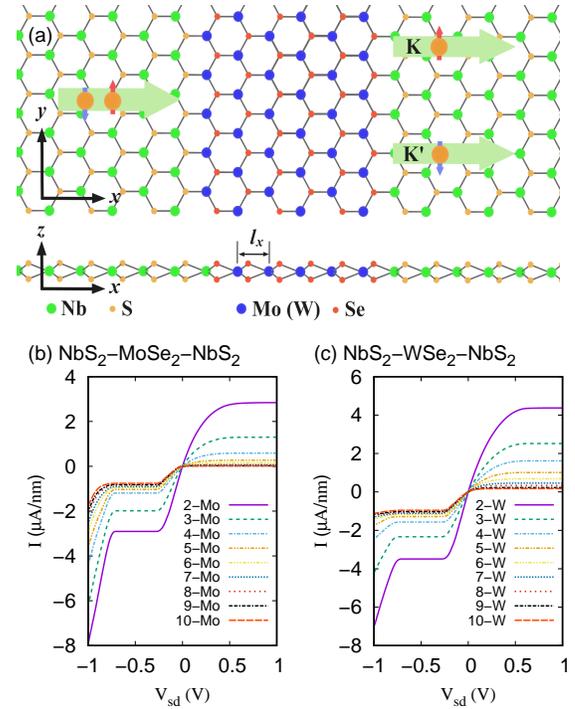}
\caption{
The schematic of lateral hetero junction of monolayer TMDCs with two interfaces in (a).  The arrows represent the incident spin-unpolarized electronic current and the transmitted electrons with spin-valley locking.
 In (b) and (c), the I-V characteristics at $T=10$K of the hetero junction with $1H$-MoSe$_2$ and WSe$_2$, respectively. The index $n$ in $n$-Mo and $n$-W represents the length of semiconducting region $nl_x$. 
 }\label{fig_schematic_MSM}
\end{center}
\end{figure}
The group VI transition-metal atoms, molybdenum and tungsten, form direct-gap semiconductors in $1H$ structure with sulfur and selenium.\cite{Mak2010,Jin2013,Zhang2013}
On the other hand,  $1H$-crystals of niobium and tantalum, group V atoms, possess metallic properties including an anomalous superconductivity in NbSe$_2$ showing fascinating phenomena\cite{Lu2015,Xi2016,Xi2017,Wang2017,Habe2019-2}.
In semiconducting $1H$-TMDCs, there is spin-valley locking of electrons in the $K$ and $K'$ valleys but the charge injection is necessary to introduce conducting electrons via a gate electrode of FET, which breaks mirror symmetry of the crystal structure.
Although the metallic $1H$-TMDCs intrinsically possess conducting channels for electrons, there are three valleys and no spin-valley correlation in the Fermi energy.
Therefore, neither metallic nor semiconducting TMDCs possess conducting electrons with the spin-valley correlation intrinsically.

In this paper, we theoretically investigate the electronic structure and transport properties of metal-semiconductor-metal lateral hetero-junctions of $1H$-TMDCs, and show the hetero-junctions intrinsically provide the spin-valley locking to conducting electrons as shown in Fig.\ \ref{fig_schematic_MSM}(a). 
Due to various electronic properties of $1H$-TMDCs, hetero structures of these monolayers have been studied for the application to electronic and optical devices in previous experimental papers\cite{Li2015,Lu2016,Son2016,Ullah2017} and theoretical papers.\cite{Sharma2014,Habe2019-1,Deilmann2020}
The lateral-hetero structure is a single layer consisting of different atomic layers as a patchwork, and that of semiconducting $1H$-TMDCs has been realized experimentally with atomically aligned interfaces.\cite{Gong2014,Duan2014,Huang2014,Chen2015,Chen2015-2,Li2015,Zhang2015,He2016,Najmzadeh2016}
We consider the lateral-hetero structure of metallic and semiconducting TMDCs for providing a fascinating electronic system with spin-valley correlation.
For investigating the electronic transport properties,  a multi-orbital tight-binding model is introduced by referring to the first-principles band structures of the pristine monolayer and periodic stripe-lattice of two TMDCs.
Moreover, the I-V characteristic of the junction is also theoretically calculated and shows that the hetero-junction provides step-like I-V characteristics, a fundamental function in conventional electronics, as shown in Figs. \ \ref{fig_schematic_MSM}(b) and \ref{fig_schematic_MSM}(c).

The remaining sections are organized as follows.
In Sec.\ \ref{sec_pristine_monolayer}, the crystal and electronic structures of monolayer NbS$_2$, MoSe$_2$, and WSe$_2$ are discussed by using first-principles calculations.
The electronic structure of s periodic lateral-hetero-structure is investigated in Sec.\ \ref{sec_lateral_hetero}.
Moreover, a theoretical model is introduced for the study of electronic transport properties.
In Sec.\ \ref{sec_transport}, the calculation method and numerical results of the electronic transport properties of metal-semiconductor-metal lateral-hetero-junctions of $1H$-TMDCs are provided.
In Sec.\ \ref{sec_discussion}, we discuss the relation between the electronic structure and the electronic transport properties, and the effect of spin-valley locking on the transport phenomena.
The conclusion is given in Sec.\ \ref{sec_conclusion}.

\section{Crystal and electronic structures of pristine monolayer crystals}\label{sec_pristine_monolayer}
\begin{figure}[htbp]
\begin{center}
 \includegraphics[width=80mm]{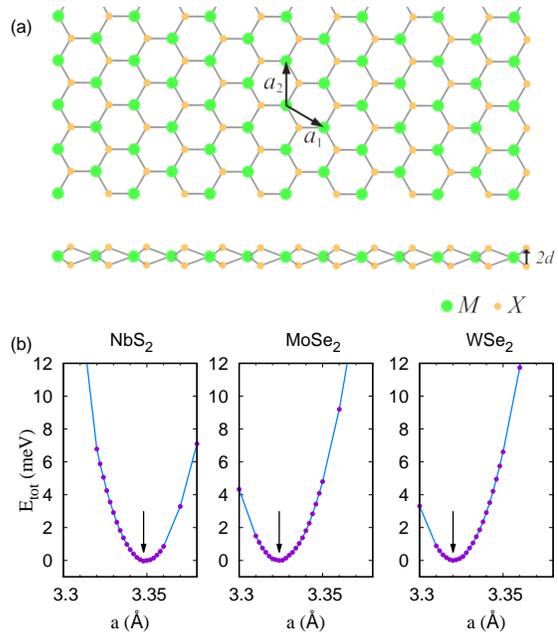}
\caption{
The schematic of lattice structure of $1H$-TMDCs in (a) and the total energy depending on the lattice constants for NbS$_2$, MoSe$_2$,  and WSe$_2$. The arrow indicates the minimum energy of stable crystal structure in (b).
The lattice constant is defined as $a=|\boldsymbol{a}_1|=|\boldsymbol{a}_2|$.
 }\label{fig_lattice_constant}
\end{center}
\end{figure}

Monolayer NbS$_2$, MoSe$_2$, and WSe$_2$ are classified into 1$H$-crystals and possess a honeycomb lattice structure consisting of three sublayers as shown in Fig.\ \ref{fig_lattice_constant}(a).
In this section, the lattice parameters and electronic structure are presented for these materials by numerical calculations using the first-principles method.
Although the electronic structure in a hetero-junction varies from that in each pristine crystal, the deviation decreases with increasing distance from the boundary of junction.
The calculations are performed by using quantum-ESPRESSO, a package of numerical codes for density functional theory (DFT)\cite{quantum-espresso}, with the projector augmented wave method including spin-orbit coupling within generalized gradient approximation.
The energy cut-off is 60Ry for the plane wave basis and 400Ry for the charge density.
The convergence criterion 10$^{-8}$ Ry is adopted for the self-consistent field calculation.

The lattice parameters of pristine crystals are estimated by comparing the total energy of electronic systems with different lattice parameters.
The numerical results for NbS$_2$, MoSe$_2$, and WSe$_2$ are presented in Fig.\ \ref{fig_lattice_constant}.
In these calculations, the lattice constant $a=|\boldsymbol{a}_1|=|\boldsymbol{a}_2|$ of a honeycomb structure is considered as a parameter and the distance $d$ between the top and bottom sublayers is simultaneously optimized by the lattice relaxation code in quantum ESPRESSO.
The numerical calculations show that the lattice constant $a$ of stable structure is almost the same among the three materials: 3.348 for NbS$_2$, 3.324 for MoSe$_2$, and 3.320 for WSe$_2$ within 1\%.
The numerical results imply the feasibility of commensurate lateral hetero-structures composed of these layered materials with an aligned interface.

\begin{figure}[htbp]
\begin{center}
 \includegraphics[width=80mm]{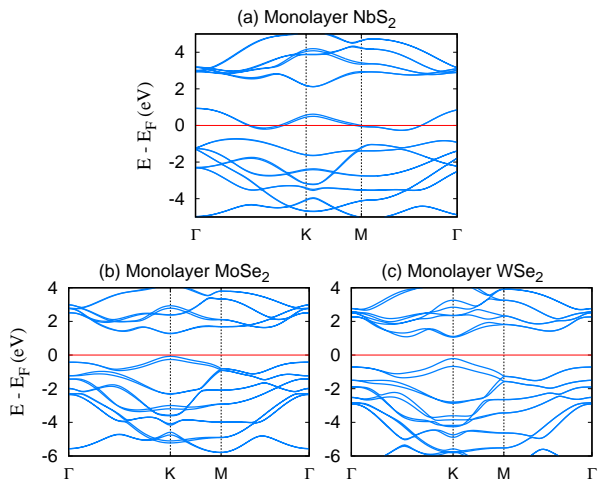}
\caption{
The electronic band structures of monolayer NbS$_2$, MoSe$_2$, and WSe$_2$. The horizontal line indicates the Fermi level.
 }\label{fig_band_pristine_layers}
\end{center}
\end{figure}
In Fig.\ \ref{fig_band_pristine_layers}, the first-principles band structures of pristine monolayers are presented by applying the optimized lattice parameters for NbS$_2$, MoSe$_2$, and WSe$_2$.
The first-principles results show the metallic character of NbS$_2$ and the semiconducting character of MoSe$_2$ and WSe$_2$.
In the semiconducting TMDCs, the top of the valence band splits at the $K$ point due to the Zeeman-type spin-orbit coupling where the heavier transition-metal atom W leads to a larger spin split.
The energy dispersion is almost the same among these crystals due to the same crystal structure.
In pristine MoSe$_2$ and WSe$_2$, the Fermi energy is inside the gap and indicates the absence of a conducting channel.
In the lateral hetero junction, however, the connecting region of these materials can possess conducting channels due to the change of the band structure and the charge transfer through the interface.

\section{Electronic structure in lateral hetero-junction}\label{sec_lateral_hetero}
In this section, electronic structures in lateral hetero-junctions are theoretically analyzed and a theoretical method is introduced for constructing a tight-binding model to investigate electronic transport properties of hetero-junctions.
The hetero-junction consists of two infinitely long metallic monolayer leads and a short semiconducting monolayer sandwiched by the leads as shown in Fig.\ \ref{fig_schematic_MSM}(a).
For the investigation, it is necessary to describe electronic states in infinitely long leads where incoming and outgoing electronic waves are traveling.
The tight-binding representation enables us to describe these electronic states in such a long system by using finite dimensional hopping matrices.
In this paper, the infinitely long junction is divided into three regions: an intermediate scattering region including two boundaries (see Fig.\ \ref{fig_band_structure_hetero}(a)) and two half infinite lead regions in a homogeneous crystal structure.
In these regions, hopping matrices are obtained by referring to different first-principles bands. 
The tight-binding model for the whole junction is constructed by connecting these hopping matrices for three regions.
The validity of the model is confirmed by comparing the first-principles band and that of the tight-binding model.

\subsection{Electronic structure in periodic junction}
\begin{figure}[htbp]
\begin{center}
 \includegraphics[width=75mm]{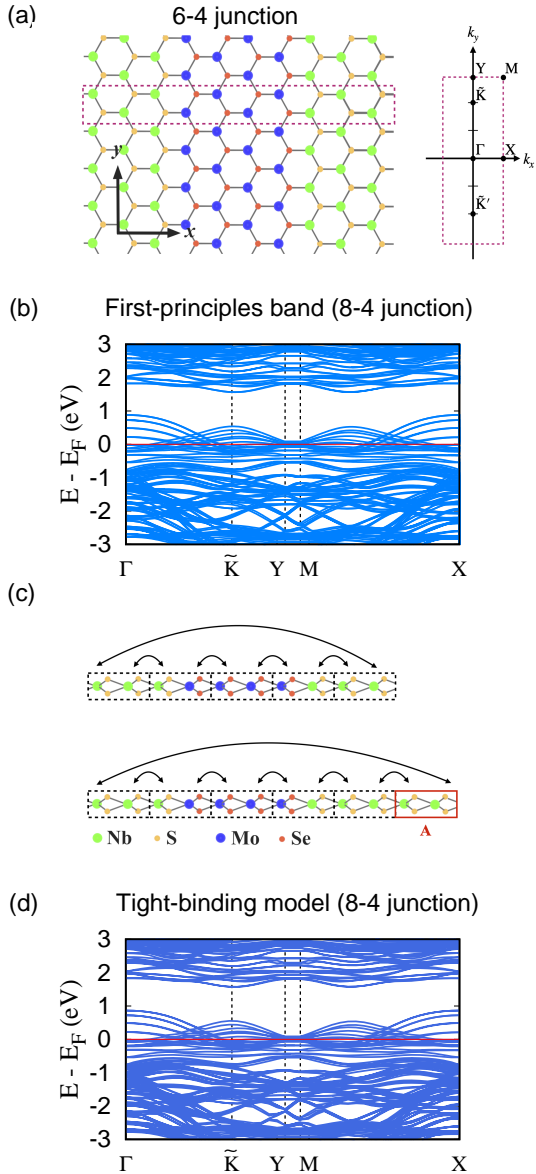} 
\caption{
The schematics and the electronic band structures of a periodic hetero-structure of NbS$_2$ and MoSe$_2$. The unit cell of the periodic structure and the first Brillouin zone are given in (a). Here, $\tilde{K}$ and $\tilde{K}'$ correspond to the $K$ and $K'$ points in the first-Brillouin zone of pristine monolayer. In (b), the first-principles band for the hetero-junction is given. In (c) and (d), the schematic of tight-binding model and the band structure by using this model are presented.
 }\label{fig_band_structure_hetero}
\end{center}
\end{figure}

The electronic structure of periodic hetero junction of NbS$_2$ and MoSe$_2$ is investigated.
The first-principles bands can be applied to the construction of a tight-binding model in the intermediate region.
The unit cell and the first Brillouin zone of a periodic junction are depicted in the left and right panels, respectively, of Fig.\ \ref{fig_band_structure_hetero}(a).
In what follows, these periodic junctions are characterized by the number of transition metal atoms in the unit cell, e.g., the $m$-$n$ hetero-junction indicates the numbers of Nb and Mo(W) atoms to be $m$ and $n$, respectively, in the unit cell.
In this paper, the zig-zag boundary structure is adopted because the experimentally observed interface is in this structure.\cite{Gong2014,Duan2014,Huang2014,Chen2015,Chen2015-2,Li2015,Zhang2015,He2016,Najmzadeh2016}
In the calculations, the lattice structure is assumed to be commensurate with a unique lattice constant, $a=3.338$\AA\ for NbS$_2$-MoSe$_2$ ($3.336$\AA\ for NbS$_2$-WSe$_2$), which provides the minimal total energy of electronic system according to the numerical results in Fig.\ \ref{fig_lattice_constant}.

In Fig.\ \ref{fig_band_structure_hetero}(b), the electronic band structure of an 8-4 periodic hetero-junction is presented as an example.
There is an energy gap from 1.0 eV to 1.8eV in the band structure.
The subbands below the gap are attributed to the valence band of MoSe$_2$ and the partial filled band of NbS$_2$.
The band structure possesses two local maxima between the $\Gamma$ and $Y$ points.
The local maximum at the $\Gamma$ point is attributed to that of the $\Gamma$ valley in the pristine TMDCs and the other local maximum at the $\tilde{K}$ corresponds to the $K$ valley. 
Thus, the electronic structure implies that the valley degree of freedom is preserved even in the hetero-junction.

\begin{figure}[htbp]
\begin{center}
 \includegraphics[width=80mm]{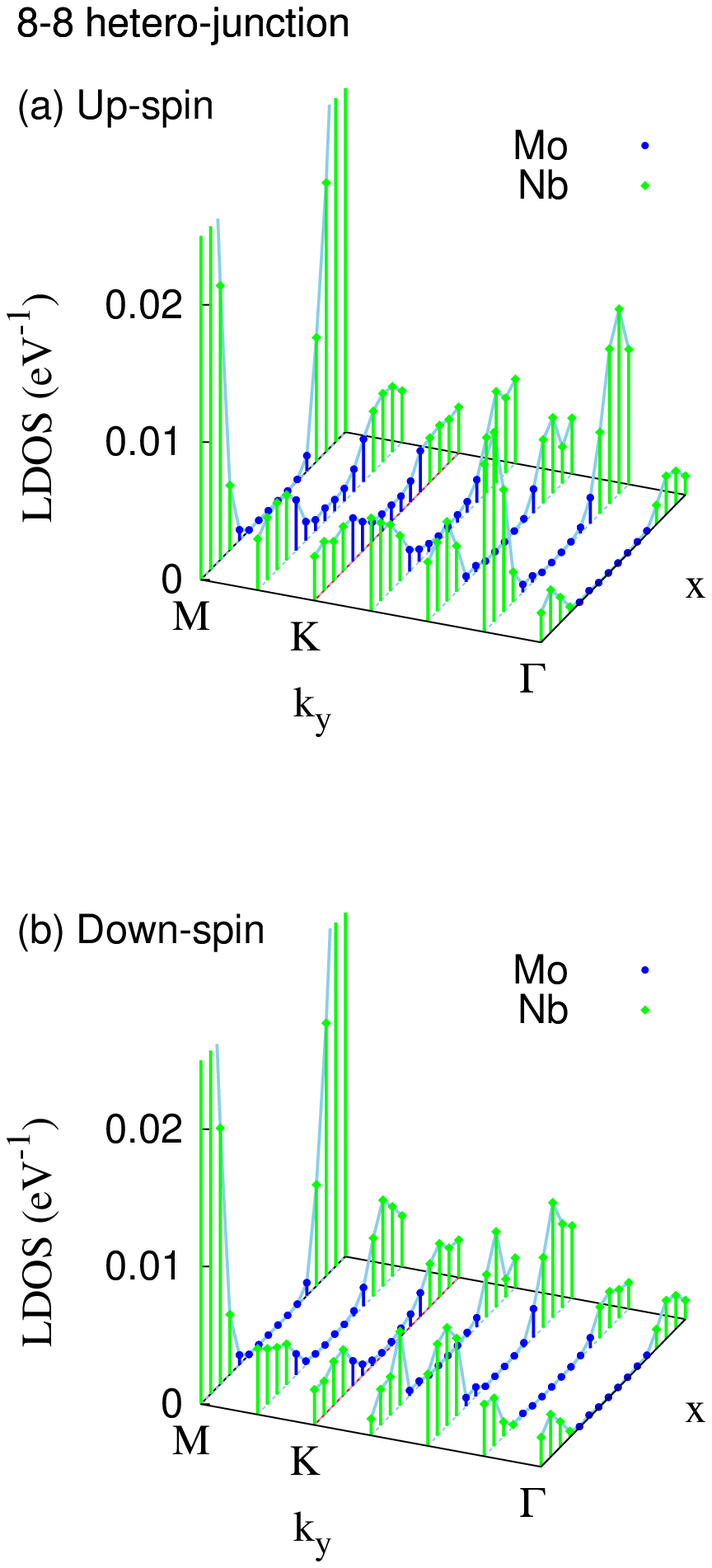}
\caption{
Spin-resolved LDOS in the Fermi level at the atomic positions of transition metal atoms for hetero-junctions of NbS$_2$ and MoSe$_2$. In the $x$-axis, each point represents the atomic position of the transition metal-atom in the junction direction. In the $k_y$-axis, $M$, $K$, and $\Gamma$ are the $k_y$ components of the wave number corresponding to high-symmetry point for the pristine $1H$-crystal.  
 }\label{fig_LDOS_periodic_junction}
\end{center}
\end{figure}
The theoretical model also reveals accumulation of charge-transfer between the metallic and semiconducting regions in the $K$ and $K'$ valleys.
The local electronic structure in the junction is investigated by using a tight-binding model for the periodic junction where the unit cell is shown in Fig.\ \ref{fig_band_structure_hetero}(a).
In the tight-binding model, the basis consists of all outer $d$-orbitals in transition-metal atoms and $p$-orbitals in chalcogen atoms within the unit cell.
The hopping integrals are calculated by using wannier90, \cite{Wannier90} a code to compute the maximally-localized Wannier orbitals and the hopping matrices between them from a first-principles band structure.
For investigating the local structure, the local density of states (LDOS) is evaluated at the atomic position of Mo and Nb atoms for several $k_y$, where the LDOS is represented by 
\begin{align}
D_{s_z,k_y}(x_j,E_F)=\sum_{n,\alpha,k_x} |\langle\psi_{n,\boldsymbol{k}}|\alpha,s_z, x_j\rangle|^2\delta(E_F-E_n),
\end{align}
with the band index $n$, the orbital index $\alpha$ in a transition-metal atom at $x_j$, and the spin index $s_z$ along the $z$-axis.
Here, $E_n$ and $|\psi_{n,\boldsymbol{k}}\rangle$ are the eigen energy and eigenvector, respectively, of the electronic state with $\boldsymbol{k}$ in the $n$-th band.
Since electronic states near the Fermi level are mainly consisting of the $d$-orbitals in transition-metal atoms, the LDOS represents the connectivity of the electronic wave around the junction in the Fermi level.

The numerical results of LDOS in a hetero-junction is presented for the up- and down-spin states in Fig.\ \ref{fig_LDOS_periodic_junction} where $M$, $K$, and $\Gamma$ represent $k_y$ corresponding to $M$, $K$, and $\Gamma$ points, respectively, in the pristine crystal.
The amplitude in the NbS$_2$ region is much larger than that in the MoSe$_2$ region.
This observation shows that the two regions of MoSe$_2$ and NbS$_2$ maintain the electronic properties of semiconductor and metal, respectively.
In the MoSe$_2$ region, the LDOS is enhanced for up-spin states around the $K$ point.
The enhancement is attributed to the charge transfer from the metallic region to the semiconducting region in the $K$ valley, where the valence top appears in the pristine monolayer.
Since the electronic structure of the down-spin electron is mirror symmetric in $k_y$, the down-spin electrons transfer in the $K'$ valley.
Therefore, electronic states are connected around the $K$ and $K'$ points with up-spin and down-spin, respectively, in the Fermi energy through the semiconducting region.

\begin{figure}[htbp]
\begin{center}
 \includegraphics[width=80mm]{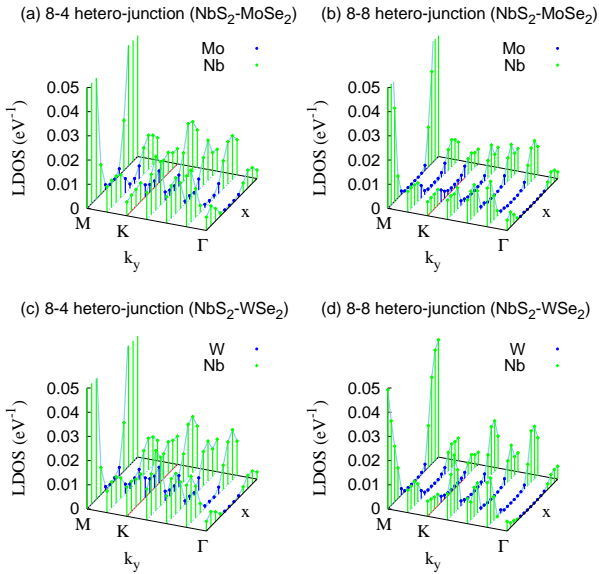}
\caption{
Total LDOS in 8-$n$ hetero-junctions for the Fermi energy. The semiconducting region consists of MoSe$_2$ for ( a) and (b), and WSe$_2$ for (c) and (d).
 }\label{fig_total_LDOS_periodic_junction}
\end{center}
\end{figure}
The LDOS varies with the atomic species and the length of semiconducting region as shown in Fig.\ \ref{fig_total_LDOS_periodic_junction}.
The numerical results show that the LDOS decreases with increase in the semiconducting region and it is enhanced within WSe$_2$.
On the other hand, it remains to be suppressed at the $\Gamma$ and $M$ points regardless of the length of the semiconducting region.
The spatial profile of the LDOS indicates that the hetero-junction restricts the electronic transmission to the $K$ and $K'$ valleys.

\begin{figure}[htbp]
\begin{center}
 \includegraphics[width=80mm]{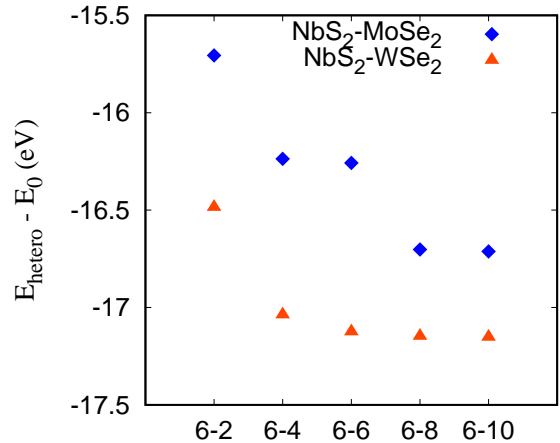}
\caption{
The energy difference between the junction and the independent flakes of NbS$_2$ and MoSe$_2$ (WSe$_2$).
 }\label{fig_forming_energy}
\end{center}
\end{figure}
Finally, the stability of the hetero-structure is also numerically confirmed by the first-principles calculations.
In Fig.\ \ref{fig_forming_energy}, the energy difference by forming the junction from the independent flakes of NbS$_2$ and MoSe$_2$ (WSe$_2$) is presented for several 6-$m$ hetero-junctions.
Since, in every $6$-$m$ junction, the total energy drops due to formation of the hetero-structure, the hetero-structure can remain stable.
The stability of the hetero-junction does not change with the length of the NbS$_2$ region.
The energy difference is enhanced with the length of the semiconducting TMDC and saturates in the 6-10 junction.
The saturation indicates that the energy change due to the addition of  MoSe$_2$ cells to the long junction is same as that to the pristine MoSe$_2$.
This implies that the local electronic structure in the middle area of the MoSe$_2$ region is similar to that of pristine MoSe$_2$.

\subsection{Tight-binding description for hetero-junctions}
A tight-binding model is constructed for the investigation of electronic transport in hetero-junctions by using hopping matrices for the periodic junction and the pristine crystal.
A hetero-junction is divided into three domains: two leads and an intermediate region including two boundaries.
The electronic states in the leads of NbS$_2$ are described by using the hopping matrices for the pristine crystal.

In the intermediate region, the tight-binding Hamiltonian for the periodic junction is applied to the description.
The unit cell is containing two transition-metal atoms and four chalcogen atoms as shown in Fig.\ \ref{fig_band_structure_hetero}(c) where it is reconstructed by dividing up that of periodic junctions shown in Fig.\ \ref{fig_band_structure_hetero}(a).
The tight-binding model is represented by
\begin{align}
H_{k_y}^{(0)}=&\sum_{n=1}^{N}\left(c^\dagger_{k_y,n}\hat{h}_{k_y,n}^\dagger c_{k_y,n}\right.\nonumber\\
&\left.+\left\{c^\dagger_{k_y,n+1}\hat{t}_{k_y,n} c_{k_y,n}+\mathrm{h.c.}\right\}\right),
\end{align}
where the index $n$ denotes the position of the cell, and $\hat{h}_{k_y,n}$ and $\hat{t}_{k_y,n}$ are the intra- and nearest-neighbor inter-cell hopping matrices, respectively.
The two cells on the both side of the center of the NbS$_2$ region are adopted as $n=1$ and $n=N$.
Here, $c_{k_y,n}$ is a state vector which consists of the annihilation operators in the spinful Wannier orbitals within the $n$-th cell, where there are ten $d$-orbitals in transition-metal atoms and twelve $p$-orbitals in chalcogen atoms for each spin.
In the periodic junction, the periodic boundary condition is given by substituting $c_{k_y,N+1}$ with $c_{k_y,1}$.

The tight-binding Hamiltonian for the infinitely long hetero-junction is represented by 
\begin{align}
H_{k_y}=&\sum_{n=-\infty}^\infty\left(c^\dagger_{k_y,n}\hat{h}_{k_y,n}^\dagger c_{k_y,n}\right.\nonumber\\
&\left.+\left\{c^\dagger_{k_y,n+1}\hat{t}_{k_y,n} c_{k_y,n}+\mathrm{h.c.}\right\}\right),\label{eq_tb_hamiltonian}
\end{align}
where the hopping matrices for the periodic junction are applied to those in $1\leq n\leq N$.
The two cells with $n=1$ and $N$ are connected to the edges of two leads of NbS$_2$ with the hopping matrix $t_{k_y,-1}=t_{k_y,N}=t_{k_y}$ for the pristine NbS$_2$.
The hopping matrices for the pristine NbS$_2$ are also applied to those in the cells for $n<1$ and $N+1\leq n$.
For the smooth connection between the intermediate and lead regions, hopping matrices at the edges of intermediate region should be similar to those in the leads.

By adopting a long NbS$_2$ region, the local electronic structure in the NbS$_2$ region of the periodic junction can be similar to that in the pristine crystal.
For example, a nearest-neighbor tight binding model for  an 8-4 periodic hetero structure is constructed by using hopping integrals computed from the band structures of a 6-4 periodic hetero-structure and pristine monolayer NbS$_2$.
The procedure is schematically depicted in Fig.\ \ref{fig_band_structure_hetero}(c).
The upper panel represents the schematic of a tight-binding model for the 6-4 periodic hetero-structure.
To construct the 8-4 model, a hopping matrix from the pristine NbS$_2$ is introduced in the cell A, the middle block in the NbS$_2$ region, as shown in the lower panel. 
The inter-cell hopping from/to the cell A is that for the 6-4 periodic hetero-structure.
In Fig.\ \ref{fig_band_structure_hetero}(d), the band structure of the 8-4 hetero-junction calculated by using this tight-binding model is presented and it is almost same as that from first-principles calculation in Fig.\ \ref{fig_band_structure_hetero}(b).
The similarity of the two band structures shows that the electronic structure in the NbS$_2$ region apart from the vicinity of the interface is almost same as those in the pristine crystal and that the model can well describe electronic states in the 8-4 hetero junction.
Therefore, the hopping matrices of the 6-$n$ periodic junction are applied to those in intermediate region of infinitely long hetero junctions.

\section{Electronic transport property of metal-semiconductor-metal junction}\label{sec_transport}
In this section, the electronic transport properties of a metal-semiconductor-metal lateral hetero junction are theoretically investigated within a two-terminal system.
The transmission probability in the hetero-junction is calculated by using the lattice Green's function method.
The theoretical calculation shows that the hetero-junction restricts the spin of transmittable electrons in each valley in the presence of a long semiconducting region.
The I-V characteristic is obtained by the Landaurer formula with the transmission probability.

\subsection{Calculation method}
The electric current is calculated by the Landauer formula,
\begin{align}
I=&\sum_{m,n}(-e)v_{x,m}[f_F(E_{m}-E_F)-f_F(E_{n}-E_F-eV)]T_{nm},\label{eq_Landauer1}
\end{align}
with a source-drain bias $V$, where $v_{x,m}$ and $f_F(\varepsilon)$ represent the velocity along the $x$ direction and the distribution function, respectively.
Here, $T_{nm}$ represents the transmission probability from the incoming state $m$ to the outgoing state $n$ in the left and right half-infinite NbS$_2$ regions, respectively.
Since electrons are scattered in a short intermediate region, the dissipation of electronic energy is ignored i.e.,  $E_m=E_n$ in the scattering process.
The atomically-aligned interface preserves the wave number in the $y$ direction.
Therefore, the Landauer formula can be rewritten by using the conductivity $\sigma(\varepsilon)$ for the electronic energy $\varepsilon=E-E_F$,
\begin{align}
I=\frac{1}{(-e)}\int{d\varepsilon}[f_F(\varepsilon)-f_F(\varepsilon-eV)]\sigma(\varepsilon),\label{eq_Landauer2}
\end{align}
with
\begin{align}
\sigma(\varepsilon)=\frac{e^2}{h}\int \frac{dk_y}{(2\pi)}\sum_{\mu,\nu}T_{\nu\mu}(\varepsilon,k_y),\label{eq_conductivity}
\end{align}
where the definition of velocity $v_{x,m}=\partial \varepsilon_m/\partial k_x$ is used.
Here, the subscripts $\mu$ and $\nu$ are the indexes to indicate the incoming and outgoing channels in the energy $\varepsilon$ at the wave number $k_y$.

The transmission probability is computed by using the Green's function method with the multi-orbital tight-binding model.\cite{Ando1991,Habe2015,Habe2016}
The Green's function describes electronic transfer from an incoming state to an outgoing state in the leads.
The wave function $c_{k_y,n}$ in leads is translational symmetric $c_{k_y,n}=\lambda c_{k_y,n-1}$ and represented by
\begin{align}
\lambda\begin{pmatrix}
c_{k_y,n}\\
c_{k_y,n-1}
\end{pmatrix}
=\begin{pmatrix}
t_{k_y,0}^{-1}(h_{k_y,0}-\varepsilon)&-t_{k_y,0}^{-1}t_\alpha^\dagger\\
1&0
\end{pmatrix}
\begin{pmatrix}
c_{k_y,n}\\
c_{k_y,n-1}
\end{pmatrix},
\end{align}
where $h_{k_y,0}$ and $t_{k_y,0}$ are the intra- and inter-cell hopping matrix in the pristine NbS$_2$.
The wave functions are classified into the right-going states, which have a positive velocity or $|\lambda|<1$, and the left-going states, which have a negative velocity or $|\lambda|>1$.
The Green's function for the junction is obtained by using the self-energy in leads,
\begin{align}
\Sigma_{0}=\hat{t}_0F_{-}^{-1},\ \ \Sigma_{N+1}=\hat{t}_0^\dagger F_{+},
\end{align}
with
\begin{align}
F_{\pm}=U_{\pm}\Lambda_{\pm}U_{\pm}^{-1},
\end{align}
where $k_y$ is omitted from the representation for simplicity.
Here, two diagonal matrices $\Lambda_{\pm}=\mathrm{diag}[\lambda_{1,\pm},\cdots,\lambda_{M,\pm}]$ and $U_\pm=[c_{1,\pm},\cdots,c_{M,\pm}]$ are introduced, where $\lambda_{j,+(-)}$ and $c_{j,+(-)}$ are the eigenvalue and the eigenvector for right(left)-going states, respectively, with the number of Wannier orbitals $M$ in the unit cell.
The Green's function $G(\varepsilon)$ for incoming electrons from the left-lead is calculated by the following procedure,
\begin{align}
G_{n,0}(\varepsilon)=&G_{n,n}(\varepsilon)\hat{t}_{n-1}G_{n-1,0}(\varepsilon),\nonumber\\
G_{n,n}(\varepsilon)=&(\varepsilon-\hat{h}_{n}-\Sigma_{n})^{-1},
\end{align}
with $\Sigma_n=\hat{t}_{n-1} G_{n-1,n-1}\hat{t}_{n-1}^\dagger$ for $1\leq n\leq N$ and the connection to the right-lead is given by
\begin{align}
G(\varepsilon)=(\varepsilon-\hat{h}_{0}-\Sigma_{N+1}-\hat{t}_{N} G_{N,N}\hat{t}_{N}^\dagger)^{-1}\hat{t}_{N}G_{N,0}(\varepsilon).
\end{align}
The transmission probability is represented by
\begin{align}
T_{\mu\nu}(\varepsilon)=\sqrt{\frac{v_\mu}{v_\nu}}[U_+^{-1}G(\varepsilon)\hat{t}_0(F_+^{-1}-F_-^{-1})U_+]_{\mu\nu},
\end{align}
where $\nu$ and $\mu$ are the indexes of incoming and outgoing channels, respectively, and it is calculated for each $k_y$.

\begin{figure}[htbp]
\begin{center}
 \includegraphics[width=85mm]{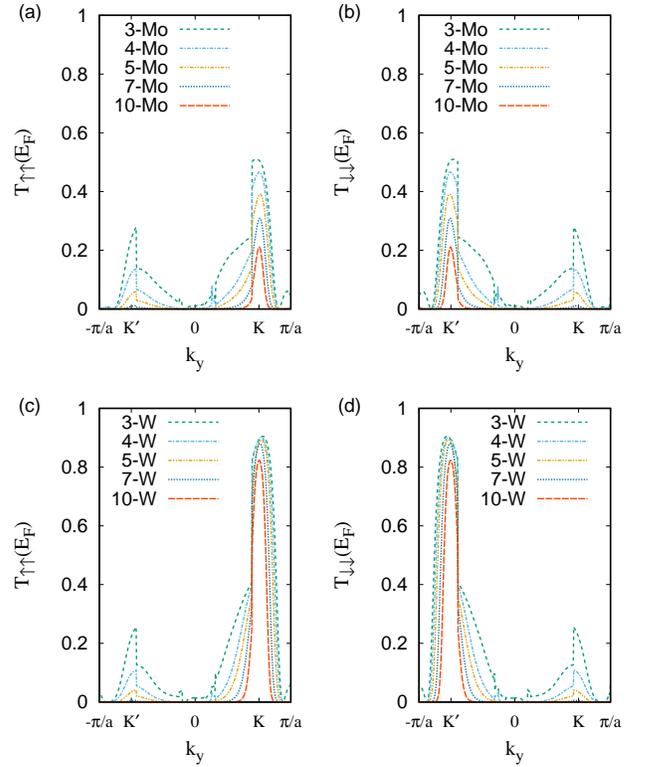}
\caption{
The electronic transmission probability through the hetero-junction of NbS$_2$-MoSe$_2$-NbS$_2$ for (a) up-spin and (b) down-spin, and NbS$_2$-WSe$_2$-NbS$_2$ for (c) up-spin and (d) down-spin. The curves represent the numerical results for the different lengths of semiconducting region. The left and right panels are the numerical results for up-spin and down spin electrons, respectively.
 }\label{fig_transmission_ky}
\end{center}
\end{figure}
\subsection{Numerical results}
Firstly, the wave number, $k_y$ of the incident electronic wave, dependence of electronic transmission probability is investigated in the Fermi energy.
In Fig.\ \ref{fig_transmission_ky}, the transmission probability is presented for several lengths of the semiconducting region.
Here, $T_{\sigma\sigma}(E_F)$ is the mean probability of all incident electrons with the spin $\sigma$ and $k_y$.
Up-spin and down-spin electrons show the inverted profiles in the $k_y$-axis because of the mirror symmetric lattice structure. 
The numerical results show electrons can transmit through the junction with high probabilities around the $K$ and $K'$ points.
Especially in the long semiconducting regions, the electronic transmission in the $K$ ($K'$) valley is permitted for up-spin (down-spin) electrons.
In the $K'$ ($K$) valley, on the other hand, the transmission probability for up-spin (down-spin) electrons rapidly decreases to zero with the length of the semiconducting region.
These observations show that the long hetero-junction enables us to produce the spin-valley locking for conducting electrons in the metallic TMDC without any exterior equipment, e.g., gate electrode or magnetic substrate, as shown in Fig.\ \ref{fig_schematic_MSM}(a).
Moreover, WSe$_2$ provides a high transmission probability over 0.8 compared with MoSe$_2$ even in the junction with the longest semiconducting region.

\begin{figure}[htbp]
\begin{center}
 \includegraphics[width=75mm]{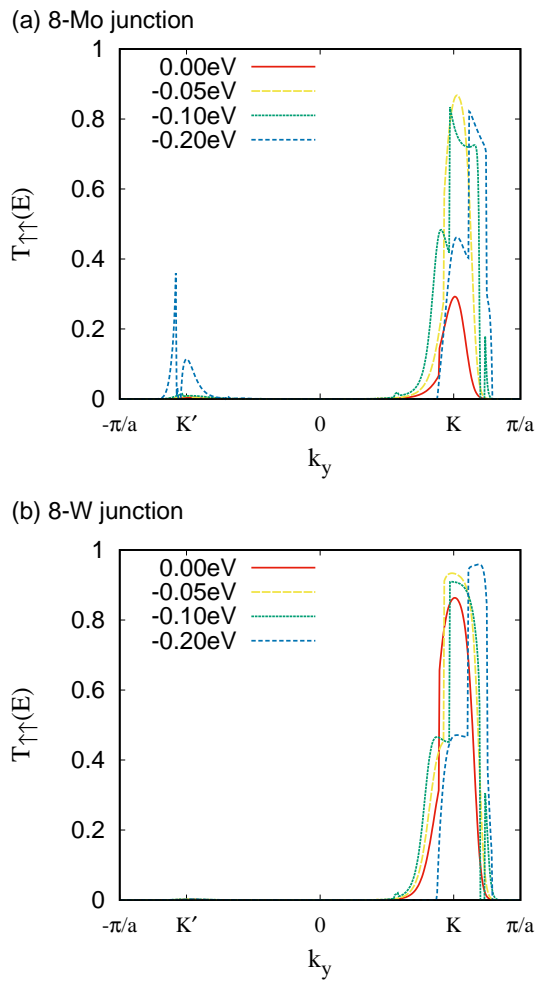}
\caption{
The mean transmission probability of up-spin electrons with several energies.  The energy is defined with respect to the Fermi energy.
 }\label{fig_transmission_E}
\end{center}
\end{figure}
The hetero-junction also provides the spin-valley locking for electrons with various energies through a long semiconducting region.
In Fig.\ \ref{fig_transmission_E},  the mean transmission probability for up-spin is presented for several energies $E<E_F$ in the hetero-junction with a long semiconducting region, where the probability for down-spin shows the inverted profile  in the $k_y$-axis.
The numerical calculations for $E>E_F$ are omitted because the electronic transmission is strongly suppressed due to the reduction of the LDOS in the semiconducting region.
Large transmission probabilities are observed only around the $K$ under all conditions.
Especially in the hetero-junction of MoSe$_2$, the decrease in $E$ enhances the transmission probability.
However, the transmission through MoSe$_2$ is also recovered in the $K'$ valley for $E-E_F<-0.20$eV even with the long semiconducting region.
These observations indicate the stability of the novel function to provide the spin-valley locking in the hetero-junction.

\begin{figure}[htbp]
\begin{center}
 \includegraphics[width=80mm]{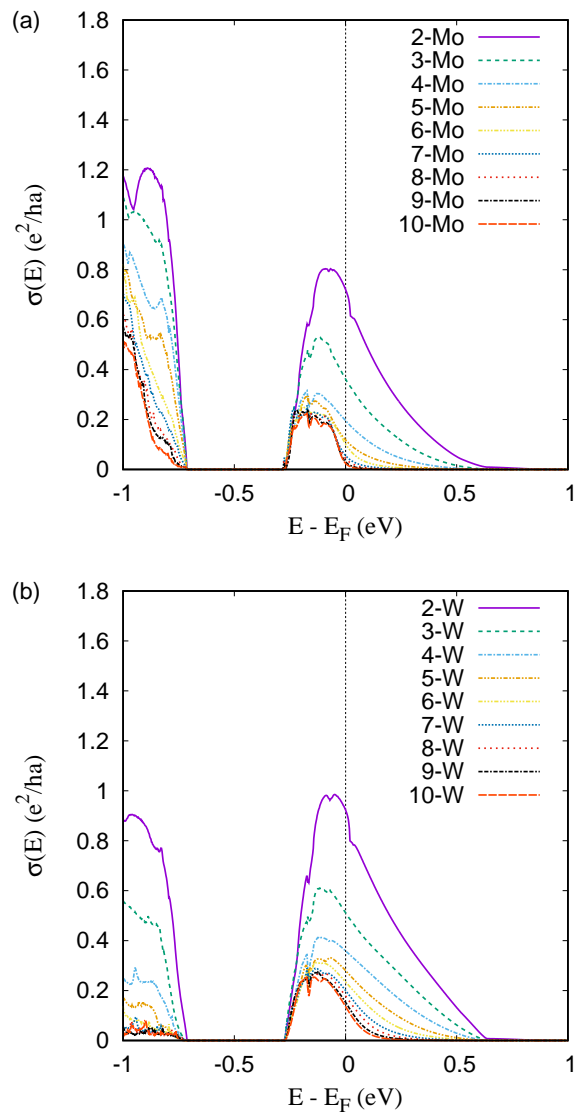}
\caption{
The electric sheet conductivity of metal-semiconductor-metal lateral hetero-junctions.  Here, $n$ of $n$-MoSe$_2$ and $n$-WSe$_2$ indicates the length of semiconducting region in units of half the box in Fig.\ \ref{fig_band_structure_hetero}(a). 
In (a) and (b), MoSe$_2$ and WSe$_2$ are adopted as the material for the semiconducting region.
 }\label{fig_conductivity_hetero}
\end{center}
\end{figure}
The electric conductivity of the hetero-junction shows characteristic features corresponding to the electronic structure.
The energy dependence of electric conductivity is presented for several hetero-junctions of MoSe$_2$ and WSe$_2$ in Fig.\ \ref{fig_conductivity_hetero}, and the I-V characteristic is shown in Fig.\ \ref{fig_schematic_MSM}(b).
The conductivity increases with decreasing the length of the semiconducting region in the whole energy region.
This observation indicates the increase of transmission probability with shortening MoSe$_2$ or WSe$_2$ regardless of energy and consistent with the variation of LDOS.
In both cases of MoSe$_2$ and WSe$_2$, there is an insulating energy region independent of the length of intermediate region and the atomic species between $-0.7$ eV and $-0.25$ eV.
This insulating region corresponds to the flat I-V characteristic in the $V<0$ and the energy gap between the partially filled band and the lower band of monolayer NbS$_2$ as shown in Fig.\ \ref{fig_band_pristine_layers}(a). 
Since this energy gap is present in NbS$_2$ and absent in NbSe$_2$\cite{Habe2020}, the insulating energy region, the flat I-V characteristic, is a unique property of NbS$_2$ lead.
The electronic transmission in the another insulating energy region in $0<E-E_F$ is attributed to the electronic properties of the intermediate region of MoSe$_2$ or WSe$_2$.
In the pristine semiconducting TMDCs, there is no conducting channel above the Fermi energy.
The small conductivity for $0<E-E_F$ is attributed to the small LDOS due to the charge transfer from the metallic region.

The I-V characteristics of metal-semiconductor-metal lateral hetero junctions show that the junction possesses an applicable electronic property in classical electronics.
In all conditions of the intermediate region, the step-like behavior is observed as a function of the bias voltage and the step height strongly depends on the length of the intermediate region.
The electric current, the step height, increases with decreasing the length but the variation is different between the positive and negative voltage regions.
In the $V<0$ region, the non-zero flow remains in junctions with long intermediate lattices of semiconducting TMDCs.
The electric current in the $0<V$ region, on the other hand, decreases to zero with the length.
The numerical results show that the rectification effect can be obtained within a monolayer hetero-junction by using a long intermediate region.
The variation of the I-V characteristic is attributed to the variation of electric conductivity $\sigma(E)$ in the two energy regions.

\section{Discussion}\label{sec_discussion}
In this section,  the discussion about the relation between the electronic transport phenomena and the electronic structure is provided.
Moreover, a possible effect of spin-valley locking on the electronic transport phenomena is discussed.
The numerical calculations show that the NbS$_2$-MoSe$_2$-NbS$_2$ and NbS$_2$-WSe$_2$-NbS$_2$ hetero junctions produce the spin-valley locking for conducting electrons.
This phenomenon is attributed to the different spin and momentum dependencies of transmission probability.
The conducting channels in the NbS$_2$ region are distributed in three valleys, $\Gamma$, K, and K', but they link to another metallic region only in the K and K' valleys through the semiconducting region.
The connection of conducting channels can be observed as non-zero LDOS in Fig.\ \ref{fig_total_LDOS_periodic_junction} and also strongly depends on the electronic spin.
This behaviour indicates that the charge transfer occurs in the $K$ valley for up-spin electrons and the $K'$ valley for down-spin electrons.
In the semiconducting $1H$-TMDCs, the highest energy occupied states appear at the $K$ and $K'$ points in the valence band and they possess the up-spin and the down-spin at the $K$ and $K'$ points, respectively, due to the Zeeman-type spin-orbit coupling.
Therefore, electrons in high energy states transfer to the NbS$_2$ region and produce the spin-polarized conducting channels around the $K$ and $K'$ points in the semiconducting region.

\begin{figure}[htbp]
\begin{center}
 \includegraphics[width=75mm]{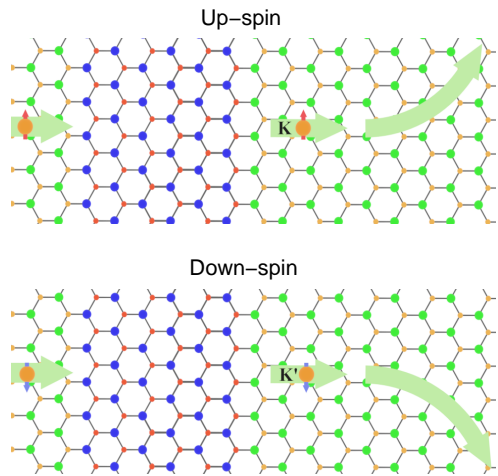}
\caption{
The schematics for transmitted electrons with each spin through the hetero-structure. The bold arrows represent the trajectory of electron.
 }\label{fig_schematic_spin}
\end{center}
\end{figure}
The spin-valley locking provides the enhancement of electronic transport phenomena related to the Berry curvature.
The electronic states in NbS$_2$ retain a similar structure of Berry curvature to that in semiconducting TMDCs, e.g., MoS$_2$.\cite{Habe2020}
In the $K$ and $K'$ valleys, the Berry curvature increases in intensity and possesses the opposite sign, and it shows the threefold symmetric sign change with a small amplitude in the $\Gamma$ valley.
Since the spin of the transmitted electron is polarized upward in the $K$ valley and downward in the $K'$ valley,  the up-spin and down-spin electrons obtain the opposite velocity in the $y$-direction attributed to the Berry curvature as shown in Fig.\ \ref{fig_schematic_spin}. 
Moreover, conducting electrons in the $\Gamma$ valley, which are less affected by the Berry curvature, are omitted from electronic transport phenomena.
Thus, for example, the hetero-junction can provide an intrinsic spin Hall system with a pure spin Hall current, which consists of the opposite flows of up-spin and down-spin electrons.
In the pristine metallic $1H$-TMDCs, on the other hand, each electronic spin possesses conducting channels in the $\Gamma$, K, and K' valleys, and it obtains different velocities in the three valleys corresponding to the sign of Berry curvature.
Therefore, the semiconducting region of a hetero-junction can drastically enhance the spin Hall effect in the metallic TMDC.

\section{Conclusion}\label{sec_conclusion}
In this paper,  we theoretically investigated the electronic transport properties of metal-semiconductor-metal lateral-hetero junctions of $1H$-TMDCs,  and revealed that the semiconducting region enables us to produce the spin-valley locking for conducting electrons in the $K$ and $K'$ valleys, and omits electrons in the $\Gamma$ valley from the electronic transport phenomena.
By using first-principles calculations, we show the stability of the hetero-junction consisting of NbS$_2$ and $M$Se$_2$ for $M$=W or Mo.
Moreover, the numerical calculations show that the charge transfer occurs between the metallic region and the semiconducting region only in the $K$ and $K'$ valleys, and it produces conducting channels in these valleys.
The long semiconducting region permits the non-zero LDOS for up-spin and down-spin electrons in the $K$ and $K'$ valleys, respectively. 
The electronic transport properties of the hetero-junction were investigated by using the lattice Green's function method with a multi-orbital tight-binding model obtained from first-principles bands.
The theoretical calculations show that the electronic transmission probability strongly depends on the spin and the wave number.
Especially in long semiconducting regions, up-spin and down-spin electrons can transmit only in the $K$ and $K'$ valleys, respectively.
Therefore, the hetero-structure produces a novel electronic system possessing conducting electrons with the intrinsic spin-valley locking.
Moreover, it is shown that the hetero-junctions are atomically thin materials with a fundamental electronic property, the step-like I-V characteristic, in classical electronics.



\bibliography{TMDC}
\end{document}